\documentclass[aps,prl,twocolumn,superscriptaddress,amsmath,amssymb,floatfix,showpacs]{revtex4}

\usepackage{graphicx}
\usepackage{ulem}

\newcommand{\FIG}[2]{\includegraphics[width=#1]{#2}}

\usepackage{bm}
\usepackage[usenames]{color}

\newcommand{\ket}[1]{\vert{#1}\rangle}

\newcommand{\KE}{-\frac{\hbar^2\nabla^2}{2m}}
\newcommand{\PE}{V}

\newcommand{\G}{\textrm{\,G}}
\newcommand{\Gcm}{\textrm{\,G/cm}}
\newcommand{\Hz}{\textrm{\,Hz}}
\newcommand{\kHz}{\textrm{\,kHz}}
\newcommand{\MHz}{\textrm{\,MHz}}
\newcommand{\GHz}{\textrm{\,GHz}}
\newcommand{\micron}{\,\mu\textrm{m}}

\newcommand{\ms}{\textrm{\,ms}}
\newcommand{\mus}{\,\mu\textrm{s}}

\newcommand{\BohrRadius}{a_0}

\newcommand{\fig}[1]{Fig.~\ref{#1}}

\newcommand{\ExpansionTime}{21.5\ms}

\newcommand{\RadialOneOne}{30.832(13)\Hz}
\newcommand{\AxialOneOne}{85.90(2)\Hz}
\newcommand{\RadialTwoOne}{31.076(13)\Hz}
\newcommand{\AxialTwoOne}{86.22(2)\Hz}

\newcommand{\gammaoneoneone}{5.4(11)\times 10^{-30}~{\rm cm}^6/{\rm s}}
\newcommand{\gammaonetwo}{0.780(19)\times 10^{-13}~{\rm cm}^3/{\rm s}}
\newcommand{\gammatwotwo}{1.194(19)\times 10^{-13}~{\rm cm}^3/{\rm s}}

\newcommand{\aoneone}{100.40\BohrRadius}
\newcommand{\aonetwo}{97.66\BohrRadius}
\newcommand{\atwotwo}{95.00\BohrRadius}

\newcommand{\NumAtoms}{3.50(5)\times 10^{5}}

\newcommand{\RF}{4.62020\MHz}
\newcommand{\MW}{6.83006800\GHz}
\newcommand{\pulselength}{0.930\ms}

\begin{document}

\title{Non-Equilibrium Dynamics and Superfluid Ring Excitations\\ in Binary Bose-Einstein Condensates}

\author{K.~M.~Mertes}
\altaffiliation[Current address: ]{Los Alamos National Laboratory, MS-J567, Los Alamos, NM 87545}

\affiliation{Department of Physics, Amherst College, Amherst,
Massachusetts 01002--5000, USA}

\author{J.~W.~Merrill}
\altaffiliation[Current address: ]{Department of Physics, Yale University, New Haven, CT 06520-8120}
\affiliation{Department of Physics, Amherst College, Amherst,
Massachusetts 01002--5000, USA}

\author{R.~Carretero-Gonz\'{a}lez}
\affiliation{Nonlinear Dynamical Systems Group\thanks{%
URL: \texttt{http://nlds.sdsu.edu/}}, Department of Mathematics and
Statistics, and Computational Science Research Center\thanks{%
URL: \texttt{http://www.csrc.sdsu.edu/}}, San Diego State University, San
Diego CA, 92182-7720, USA}

\author{D.~J.~Frantzeskakis}
\affiliation{Department of Physics, University of Athens,
Panepistimiopolis, Zografos, Athens 157 84, Greece}

\author{P.~G.~Kevrekidis}
\affiliation{Department of Mathematics and Statistics, University
of Massachusetts,  Amherst, Massachusetts 01003-4515, USA}

\author{D.~S.~Hall}
\affiliation{Department of Physics, Amherst College, Amherst,
Massachusetts 01002--5000, USA}

\pacs{03.75.-b, 03.75.Dg, 03.75.Mn}

\date{To appear in {\em Phys.~Rev.~Lett.}, 2007}

\begin{abstract}
We revisit a classic study
[D. S. Hall {\it et al.}, Phys. Rev. Lett. {\bf 81}, 1539 (1998)]
of interpenetrating Bose-Einstein condensates in the hyperfine states
$\ket{F = 1, m_f = -1}\equiv\ket{1}$ and $\ket{F = 2, m_f = +1}\equiv\ket{2}$
of ${}^{87}$Rb and observe striking new non-equilibrium component separation
dynamics in the form of oscillating ring-like structures. The process of
component separation is not significantly damped, a finding that also
contrasts sharply with earlier experimental work, allowing a clean first
look at a collective excitation of a binary superfluid. We further
demonstrate extraordinary quantitative agreement between theoretical
and experimental results using a multi-component mean-field model with
key additional features: the inclusion of atomic losses and the careful
characterization of trap potentials (at the level of a fraction of a percent).
\end{abstract}

\maketitle

Binary mixtures of Bose-Einstein condensates (BECs)~\cite{Myatt1997a,Hall1998a,Stamper-Kurn1998b} display rich phase separation behavior that is driven primarily by the nonlinear interactions between the different atomic species or states that make up the condensates. This places them at a remarkable interface between atomic physics, nonlinear and wave physics, and non-equilibrium statistical physics. The first experiment to directly probe the dynamics of binary BECs~\cite{Hall1998a} found complex motion that tended to preserve the total density but quickly damped to a stationary state with non-negligible component overlap. Other experiments have since explored various regimes. For example, partially condensed gases of $^{87}$Rb exhibited dramatic transient non-diffusive spin polarizations reminiscent of longitudinal spin waves~\cite{Lewandowski2002a}. Rotating binary BECs were observed to evolved from an initially coincident triangular vortex lattice through a turbulent regime into an interlaced square vortex lattice~\cite{Schweikhard2004a}. In optical traps, BECs consisting of atoms in different Zeeman levels of $^{23}$Na formed striated magnetic domains~\cite{Miesner1999a,Stenger1999a} between which the spin could tunnel~\cite{Stamper-Kurn1999b}. In $^{87}$Rb, different Zeeman levels in a quenched condensate formed ferromagnetic domains of variable size with interesting topological features~\cite{Sadler2006a}.

In understanding component separation, theoretical investigations~\cite{Ho1996a,Esry1997a,Law1997a,Ao1998a} have demonstrated that the stationary state of a BEC mixture depends critically on the intra- and inter-species scattering lengths, as does its stability against excitations. On the other hand, theoretical studies of static and dynamic
properties~\cite{Goldstein1997a,Busch1997a,Graham1998b,Esry1998a,Pu1998b}, such as the excitation spectrum and the nature of low-frequency simultaneous collective excitations, remain unverified experimentally. In this paper we begin to fill in the gaps between theory, numerical simulation, and experiment.

\begin{figure*}[ht!]
\FIG{1.0\textwidth}{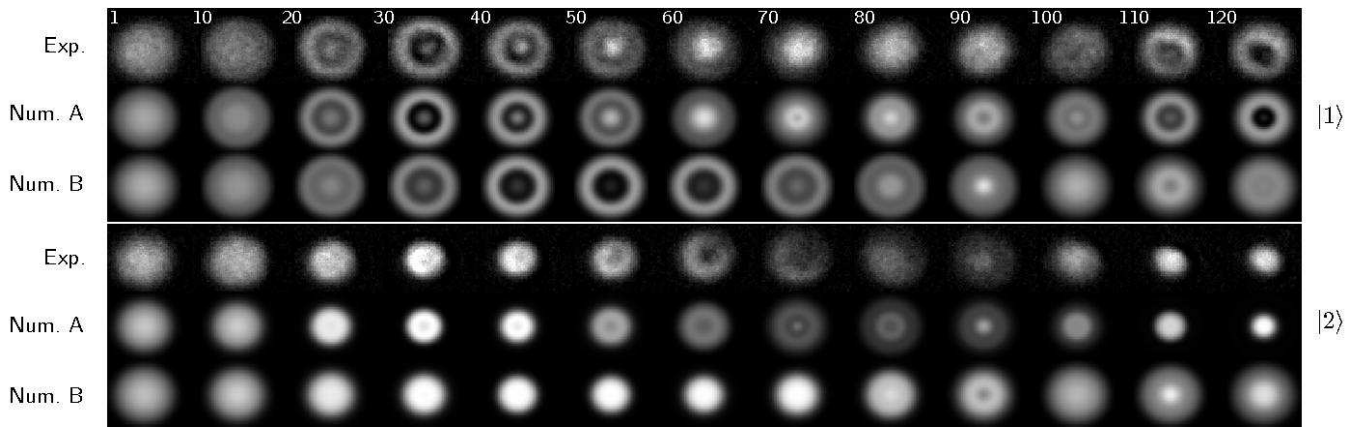}
\caption{\label{timeEvolution}
Top view of a time-sequence of experimental and numerical density profiles for $N=\NumAtoms$ $^{87}$Rb atoms with equal populations in the $\ket{1}$ and $\ket{2}$ states. The first row shows the measured density profiles for the $\ket{1}$ atoms, while the second and third rows give numerical results including losses and different trap frequencies (Num.~A) and without those additional model features (Num.~B). A similar arrangement is given for the $\ket{2}$ atoms in the fourth, fifth, and sixth rows. The field of view in all pictures is approximately $100\micron$ on a side. The evolution time (in ms) for each column is indicated in the top row.}
\end{figure*}

Our experimental approach follows closely that of the first interpenetrating condensate experiment at JILA~\cite{Hall1998a}. We begin with a single condensate of $^{87}$Rb in the $\ket{1}$ state in a time-averaged orbiting potential (TOP) magnetic trap~\cite{Petrich1995a}. The magnetic quadrupole gradient of $144(1)\Gcm$ (axial) and rotating magnetic bias field of $8.32(2)\G$ were adjusted to confine both hyperfine states with coincident potential minima~\cite{Hall1998a}. We used RF spectroscopy to establish that the magnitude of the rotating bias field varies by no more than $\pm 0.3\%$ over a full rotation --- thereby controlling one of the dominant contributions to the cylindrical asymmetry of the trap. The radial [axial] trapping frequencies were measured to be $\RadialOneOne~[\AxialOneOne]$\ for the $\ket{1}$ state and $\RadialTwoOne~[\AxialTwoOne]$ for the $\ket{2}$ state --- a difference of about $0.8\%~[0.4\%]$ that agrees with a calculation that includes the effect of quadratic Zeeman terms.

A $\pulselength$ two-photon ($\nu_{\textrm{rf}}=\RF$, $\nu_{\textrm{mw}}=\MW$) $\pi/2$-pulse~\cite{Matthews1998a} with approximately $500\kHz$ intermediate-state detuning then drives $50(2)\%$ of the atoms from the $\ket{1}$ state to the $\ket{2}$ state. After the pulse ends, the system is allowed to evolve for a time $t$. During this evolution, a weak radiofrequency ``shield'' applied at $6.3\MHz$ reduces the influence of trapped thermal atoms produced by inelastic losses. The trap is turned off by reducing the magnetic quadrupole to zero in $\sim 200\mus$; the BEC expands as it falls. Finally, $\ExpansionTime$ after releasing the atoms from the trap, we reduce the bias field to $1.0\G$ and image~\cite{noteImage} the atoms in either state with one of two cameras, the first oriented to view the atoms from the top (along the direction of gravity and axis of symmetry), and the second along a perpendicular axis. The entire procedure is repeated for various evolution times $t$, each with the same number of atoms $N$, resulting in a series of post-expansion top- and side-view images depicting the evolution of each component of the condensate.

After the transfer, the system can be described theoretically by a pair of coupled nonlinear partial differential equations of the Gross-Pitaevskii (GP) type \cite{Goldstein1997a,Busch1997a,Graham1998b,Esry1998a,Pu1998b}:
\begin{eqnarray*}
i\hbar\frac{\partial\psi_1}{\partial t}&
\hskip-2pt=\hskip-2pt&\left[\KE+\PE_1+g_{11}\vert\psi_1\vert^2+g_{12}\vert\psi_2\vert^2-i\Gamma_{1}\right]\psi_1,
\\
i\hbar\frac{\partial\psi_2}{\partial t}&
\hskip-2pt=\hskip-2pt&\left[\KE+\PE_2+g_{22}\vert\psi_2\vert^2+g_{12}\vert\psi_1\vert^2-i\Gamma_{2}\right]\psi_2,
\end{eqnarray*}
where $g_{ij}=4\pi\hbar^2 a_{ij}/m$ are the coupling constants for elastic interactions between atoms in states $\ket{i}$ and $\ket{j}$ with scattering lengths $a_{ij}$, and $\PE_i$ represents the trapping potential for species $\ket{i}$. The key features that distinguish the present model from earlier ones are (i) the different trap frequencies for different species, and (ii) the explicit inclusion and accurate quantitative characterization of the loss rates $\Gamma_i$. Here, $\Gamma_{1}=\frac{\hbar}{2}(\gamma_{111}\vert\psi_1\vert^4+\gamma_{12}\vert\psi_2\vert^2)$ and $\Gamma_{2}=\frac{\hbar}{2}(\gamma_{12}\vert\psi_1\vert^2+\gamma_{22}\vert\psi_2\vert^2)$ represent losses for species $\ket{1}$ and $\ket{2}$, respectively. Physically, three-body recombination losses dominate in a pure $\ket{1}$ BEC (decay constant $\gamma_{111}=\gammaoneoneone$~\cite{Burt1997a}), but two-body inelastic collisions dominate in mixed and pure $\ket{2}$ BECs (decay constants $\gamma_{12}=\gammaonetwo$ and $\gamma_{22}=\gammatwotwo$, values we report here for the first time, as obtained from loss data with mixed and pure condensates, respectively). Other loss processes are of limited experimental significance in this particular system and thus are not included in the GP equations.

The binary condensate begins in a highly non-equilibrium state as a
result of the transfer of atoms and small differences in the inter-
and intra-species scattering lengths. One would then naturally expect
the system to evolve in the direction of its stationary ground state,
which consists of a ball of $\ket{2}$ atoms surrounded by a shell of
$\ket{1}$ atoms~\cite{Pu1998a}.
Such evolution was observed in Ref.~\cite{Hall1998a}, in which the system apparently arrived at its ground state without exhibiting any oscillatory behavior. In our experiment the motion is {\em not} rapidly damped~\cite{noteDamping2}, in marked contrast to the earlier results~\cite{noteDamping}. A typical time sequence is shown in \fig{timeEvolution}, which we present alongside the results of the corresponding numerical simulations~\cite{noteNumerics} with and without the key model additions. The superb agreement of the improved model (Num. A) with the present experiment emphasizes one of our main results: the motional damping observed in Ref.~\cite{Hall1998a} is not intrinsic to the component separation process.

The shortcomings of the unimproved model (Num. B) are clearly visible in \fig{timeEvolution}, as it exhibits ring structures when the experiment does not (e.g., at 60~ms) and does not exhibit rings when the experiment does (e.g., at 120~ms). We find the modified timing of these features in the improved model is due primarily to the different trap frequencies experienced by states $\ket{1}$ and $\ket{2}$. 
Both of the key model enhancements also contribute to more subtle changes 
in the coetaneous atomic density distributions.

\begin{figure}[ht]
\FIG{8.58cm}{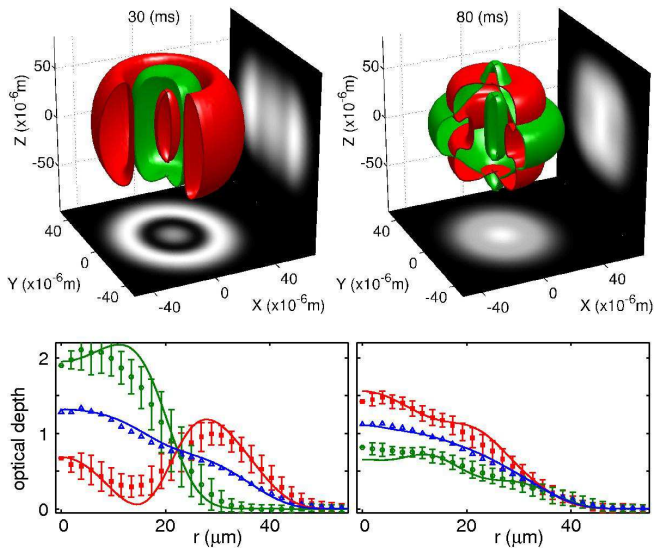}
\caption{\label{tomatoes}
(color)
Top panels: 3D renderings of the density distributions in a binary BEC computed from the model at 30 and 80~ms. Each component is depicted by a contour slice at about half of its corresponding maximal density. Red and green surfaces correspond to components $\ket{1}$ and $\ket{2}$ respectively. The bottom (side) projection corresponds to the $z$- ($x$-) integrated density for the $\ket{1}$ component in our model. Bottom panels: radial profiles of the density distributions at 30 and 80 ms. Symbols (squares for $\ket{1}$ and circles for $\ket{2}$) correspond to the experimental data, which are averaged over the azimuthal angle. Solid lines depict the model results. The middle curve depicts half of the total density, which is nearly preserved during the time evolution~\protect\cite{Hall1998a}.}
\end{figure}

The ring patterns in state $\ket{1}$ at $t=30$ and $120\ms$,
as well as that of state $\ket{2}$ at $t=70\ms$, constitute
the most striking manifestation of the dynamical evolution
and phase separation of the hyperfine state components of
our system. Taking one period to be the time between the two
ring-like patterns appearing at $30\ms$ and $120\ms$, we
determine an oscillation frequency $\nu\approx 11\Hz$ --- much
lower than the radial trap frequency, $\nu_{r}\approx 31\Hz$.
This behavior is quite different from that of single-species
BECs where excitation frequencies are always greater than the
trap frequency~\cite{Busch1997a,Pu1998a}, and further emphasizes
the essentially two-component nature of the excitation~\cite{Pu1998b}.
We also observe that the two species oscillate out of phase so that
the total density remains nearly constant in time~\cite{Hall1998a}.
A second feature also emerges from inspection of the ring patterns:
at $30\ms$, the distribution of $\ket{1}$ atoms has a central peak
(``bull's eye'') which is less pronounced in the recurrence at $120\ms$.
As the system evolves it does not precisely revisit its earlier
configuration due to atomic losses and incommensurate radial and
axial trap frequencies.


\begin{figure}[ht]
\FIG{8.58cm}{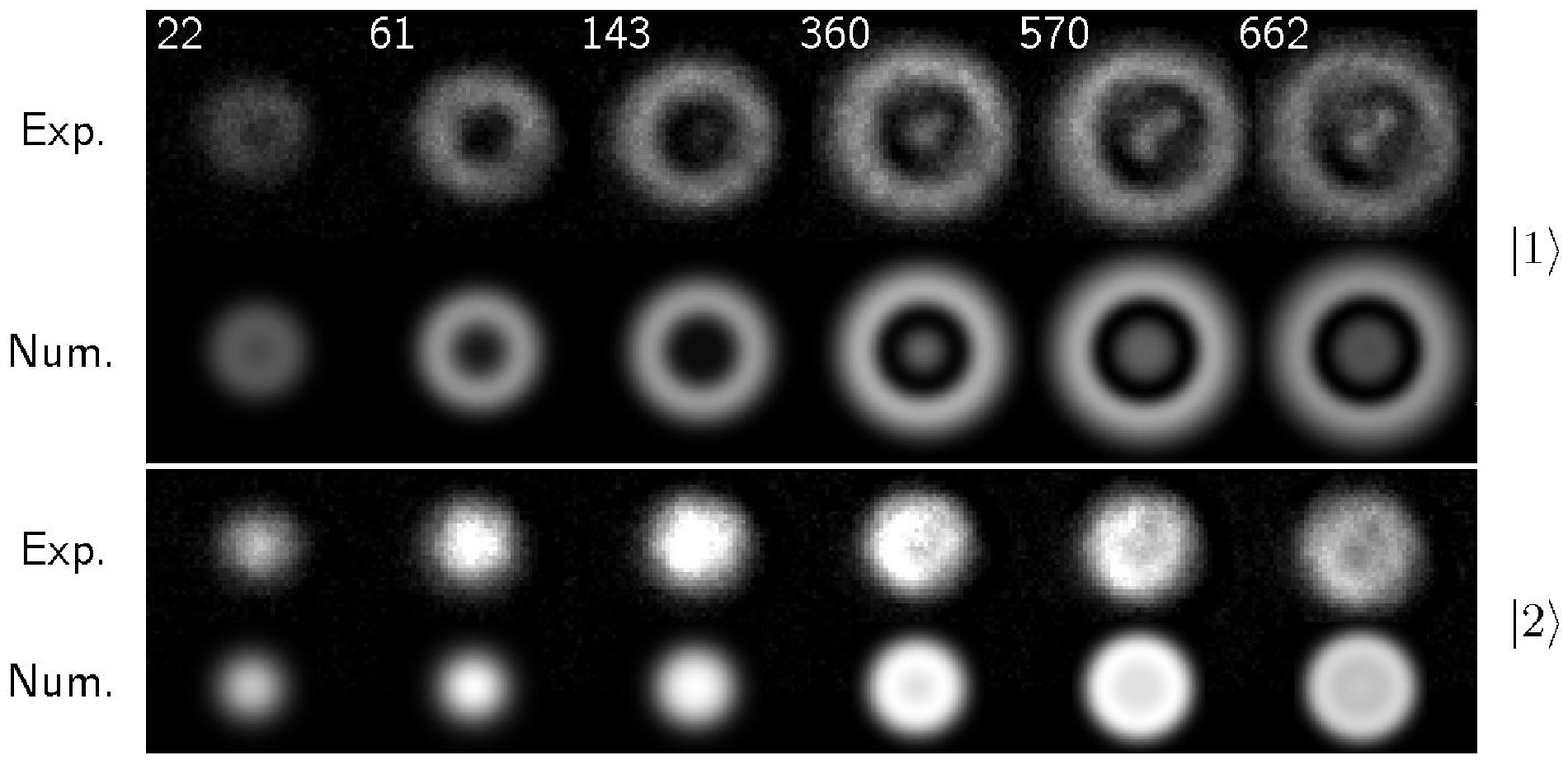}\\[1.0ex]
\FIG{8.58cm}{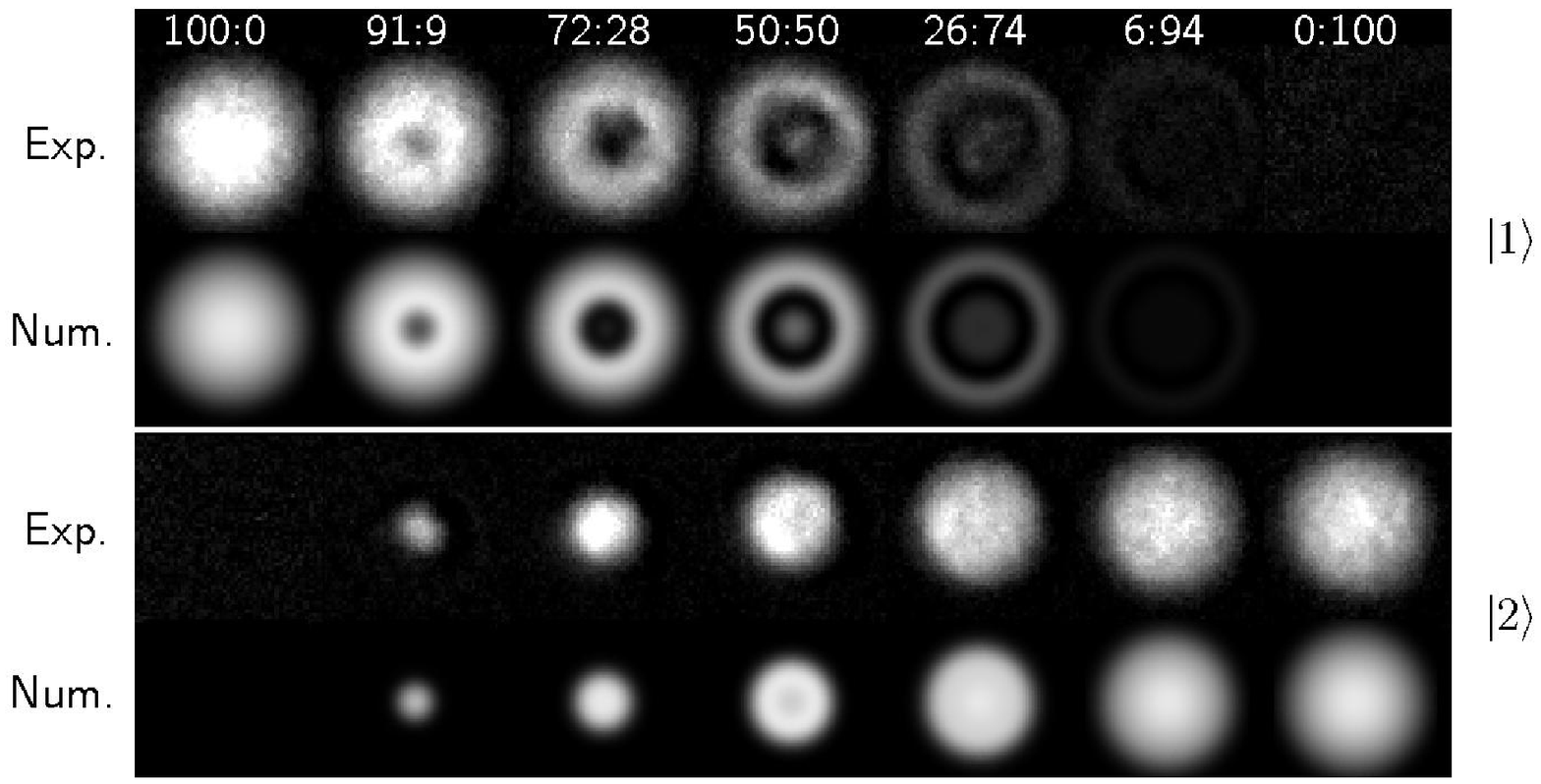}
\caption{\label{totalNumber}
Top panels: Experimental and numerical density profiles of a 50:50 mix of $\ket{1}$ and $\ket{2}$ atoms at $30~\ms$ for different $N$ as indicated (in thousands of atoms). Different experimental scalings~\cite{noteImage} are used to avoid saturating the images.
Bottom panels: same as the preceding for different relative numbers of atoms as indicated by the $\ket{1}$:$\ket{2}$ ratio. The atom number is fixed at $N=\NumAtoms$, as in Fig.~\protect\ref{timeEvolution}.}
\end{figure}

We used radial moments (such as $I_1=\int r |\psi|^2 dV$) to examine the consistency of the best theoretical values of the scattering lengths~\cite{KokkelmansPC} by comparing numerically computed and experimentally measured density profiles. Through a detailed analysis of the periodicity of $I_1(t)$, which represents the collective excitation leading to the recurrence of the ring waveform in \fig{timeEvolution}, we are able to show that the value $a_{22}=\atwotwo$ agrees well with the experimental data. By using the previously computed values $a_{11}=\aoneone$ and $a_{12}=\aonetwo$~\cite{KokkelmansPC}, we have obtained remarkable agreement between the experimental and the numerical results in all the cases considered up to 130~ms. At longer times, effects of the imperfect cylindrical symmetry of the trap become increasingly apparent in the relative position of the two condensates and the crispness of the recurring ring patterns.

Our model can be used to infer the complex spatial configurations arising from the nonlinear interactions between the two components. We show in Fig.~\ref{tomatoes} a 3D rendering of the spatial distribution of the densities for both species based upon the numerical results, which reveals interesting aspects of the interaction effects such as the location of the rings in the $z$-direction~\cite{noteMovies}. The modeled density distributions provide a richer view of the separation phenomena and are more subtle than might be suspected from an analysis of the experimental data, which consist only of integrated planar projections of the atomic density distributions.

We have also explored the dependence of the period and shape of the ring-like structures on the total initial number of atoms and the relative atom fraction in each of the two spin states. Our observations at $t=30$~ms as we vary $N$ and the relative number are shown in \fig{totalNumber}. We find that although the precise density distribution at any particular time depends sensitively on both of these variables, the periodicity of the patterns and the quality of agreement between the experimentally obtained and the numerically computed profiles do not.

The observed ring patterns are reminiscent of radial excitations, and the further we drive the system out of equilibrium the more rings are formed. In the top panels of \fig{totalNumber}, the number of rings increases as the mean-field (nonlinear) interaction energy of the system increases. Similarly, in the bottom panels of \fig{totalNumber}, the number of rings increases as more atoms are transferred from $\ket{1}$ to $\ket{2}$. We understand this behavior by noting that the excess energy after transfer, as determined by differences in scattering lengths and trapping potentials, places the system far from its ground-state equilibrium and consequently induces a long oscillatory journey through the system's configuration space. Stronger interactions, larger transfer ratios, and greater differences in the trapping potentials should therefore produce more dramatic dynamics. The experiment demonstrates that the first two conditions lead to the most pronounced ring-like structures; the last will be explored elsewhere.

In conclusion, we have observed for the first time a largely undamped two-component collective excitation in which each component forms striking ring-like patterns that recur on a timescale much longer than the radial period. Numerical solutions to a theoretical model based on a modified pair of coupled Gross-Pitaevskii equations, along with careful experimental characterization of loss rates and trapping potentials, accurately and quantitatively describe the dynamics of the system, and help to visualize aspects of the complicated component separation behavior that are difficult to observe directly.
We anticipate the possibility of assuming the accuracy of the model and using experimental data to extract scattering length ratios from the periodic behavior of the density distributions, much as in Ref.~\cite{Matthews1998a}. This technique would be sensitive to the interspecies scattering length, of crucial importance to the question of the miscibility of two atomic species that depends on the value of the determinant $D=a_{11} a_{22}-a_{12}^2$~\cite{Ho1996a,Esry1997a,Law1997a,Ao1998a}. A further extension involves self-consistently including beyond mean-field effects, such as thermal excitations, in order to match the corresponding component separation experiments at finite temperature. These directions are subjects of future work.

This effort was supported by the NSF through grants PHY-0140207, PHY-0457042,
DMS-0204585, DMS-0349023, DMS-0505663 and DMS-0619492.


\begin{thebibliography}{28}
\bibitem[{not({\natexlab{a}})}]{noteImage}
\bibinfo{note}{For $\ket{1}$ state detection, a resonant laser pulse removes
  the $\ket{2}$ atoms; then a fraction (typically $12\%$) of the $\ket{1}$
  atoms are transferred to the $\ket{2}$ state with a short two-photon pulse.
  The $\ket{2}$ atoms are subsequently imaged on resonance, giving an
  unsaturated, selectively scaled density profile for the $\ket{1}$ condensate.
  For $\ket{2}$ state detection this procedure is preceded by a
  population-inverting two-photon $\pi$-pulse. This new imaging method avoids
  difficulties associated with lensing and saturation.}
\bibitem[{not({\natexlab{b}})}]{noteDamping2}
\bibinfo{note}{Atomic losses become significant on a timescale $\gtrsim
  100$~ms and should not affect the oscillatory behavior significantly at
  shorter times. Other possible damping mechanisms, such as Beliaev damping and
  Landau damping, may be inhibited by the low frequency of the excitation and
  by the low sample temperature, respectively.}
\bibitem[{not({\natexlab{d}})}]{noteNumerics}
\bibinfo{note}{We numerically integrate the coupled GP equations using a
  standard finite-difference in space, fourth-order in time scheme with the
  equations expressed in cylindrical coordinates. From the symmetry of the
  system the equations become effectively two-dimensional. We adaptively
  increase the number of mesh points to keep up with the growing domain after
  releasing the condensate from the trap.}
\end{thebibliography}

\end{document}